# Skyrmions-based logic gates in one single nanotrack completely reconstructed via chirality barrier

Dongxing Yu[1], Hongxin Yang[1,4,*], Mairbek Chshiev[2,5], Albert Fert[3]

[1] Quantum Functional Materials Laboratory, Ningbo Institute of Materials Technology and Engineering, Chinese Academy of Sciences, Ningbo 315201, China
[2] Université Grenoble Alpes, CEA, CNRS, Spintec, 38000 Grenoble, France
[3] Unité Mixte de Physique, CNRS, Thales, Université Paris-Sud, Université Paris-Saclay, Palaiseau 91767, France
[4] Center of Materials Science and Optoelectronics Engineering, University of Chinese Academy of Sciences, Beijing 100049, China
[5] Institut Universitaire de France (IUF), 75231, Paris, France

**ABSTRACT:** Logic gates based on magnetic elements are promising candidates for the logic-in-memory applications with nonvolatile data retention, near-zero leakage and scalability. In such spin-based logic device, however, the multi-strip structure and fewer functions are obstacles to improving integration and reducing energy consumption. Here we propose a skyrmions-based single-nanotrack logic family including AND, OR, NOT, NAND, NOR, XOR, and XNOR which can be implemented and reconstructed by building and switching Dzyaloshinskii-Moriya interaction (DMI) chirality barrier on a racetrack memory. Besides the pinning effect of DMI chirality barrier on skyrmions, the annihilation, fusion and shunting of two skyrmions with opposite chirality are also achieved and demonstrated via local reversal of DMI, which are necessary for the design of engineer programmable logic nanotrack, transistor and complementary racetrack memory.

*Email: hongxin.yang@nimte.ac.cn

# INTRODUCTION

Magnetic skyrmions are non-trivial spin structures with topological protection [1-3] that exhibit many desirable features, such as nanoscale size, remarkable stability, and low driving threshold current [4-10]. These characteristics make them suitable for the design of nonvolatile, low-power spin-logic gates which can be integrated with memory [11-24]. Since the realization of two or more spin-based logic functions in such architectures usually requires the combination of multiple strips or the cascade of simple operations, the multifunctional spin-logic devices have more room for simplification to improve integration and reduce energy consumption of integrated circuits. A possibility of switching Dzyaloshinskii-Moriya interaction (DMI), a crucial ingredient for the formation of exotic chiral magnetic states [25,26], represents a very attractive direction and an important challenge [27-32]. By reconstructing different chirality barriers (i.e. the energy boundary between states of opposite DMI chirality proposed in this work) using electric field pulse, one can effectively control the dynamics of skyrmions, thereby establishing an additional degree of manipulation for the programmable skyrmion-based logic devices.

Logic gates usually consist of two inputs and one output. Considering the very simple structure of one single nanotrack, we propose to create magnetic skyrmions at both ends of it via magnetic tunnel junction (input-MTJ) as inputs [19,23,33], and set up an output-MTJ in the middle area to detect presence or absence of magnetic skyrmion as output. The design of reconfigurable single-nanotrack logic gates needs to solve two problems. The first one is controlling the relative motion of two magnetic skyrmions serving as the inputs in order to make they meet at the output. Secondly, logic gates with various functions need different outputs response to the same input, which require the ability to switch between different skyrmion dynamics by external stimuli to realize the reconstruction of logic gates. The emergence of DMI chirality barrier considered in this work can effectively solve these problems. For instance, it enables skyrmions with opposite chirality to meet and interact by locally reversing the DMI chirality in a nanotrack as shown in Fig. 1a and Fig. 2a. The distribution and number of chirality barriers can be changed by electric field pulse or other methods that can control DMI [27-31,34-36], and the resulting various magnetic skyrmion dynamic effects will allow different logic operations to be performed efficiently within one single nanotrack.

In this work, besides the pinning effect of DMI chirality barrier on magnetic skyrmions, the annihilation, fusion and shunting of the skyrmion-skyrmion pair with opposite chirality are also realized by switching the type of the barrier (single and double barriers). Then, the completely reconfigurable logic family including AND, OR, NOT, NAND, NOR, XOR, and XNOR are implemented based on the proposed skyrmion dynamics in one single nanotrack. Such a simple single-nanotrack skyrmion-based logic device can realize all logic calculation functions and switch from one operation to another without the superposition of logic gates, which is more suitable for the device integration. By controlling the local chirality of DMI in a racetrack memory, we also demonstrate the application of DMI chirality barrier in the skyrmion transistor and the reset process of skyrmion bits.

**RESULTS AND DISCUSSION**

**Model and skyrmion motion phase diagram**

We consider a CrN multiferroic monolayer as an example but not limited to it. The energy density $E$ of our magnetic system can be written as

$$E = A(\nabla \boldsymbol{m})^2 - K_{\text{uz}}(m_z)^2 - \frac{1}{2}\mu_0 M_s \boldsymbol{m} \cdot \boldsymbol{H}_\text{d} + D(m_z(\nabla \cdot \boldsymbol{m}) - (\boldsymbol{m} \cdot \nabla)m_z),$$

with each term representing exchange coupling, perpendicular magnetic anisotropy (PMA), demagnetization, and DMI energy terms, respectively. As shown in our recent work [31], the PMA here is strong enough, so the above energy density $E$ allows the existence of a ~10 nm isolated metastable skyrmion without external magnetic field. The topological property of a skyrmion can be described by the topological charge $Q = 1/(4\pi) \int \boldsymbol{m} \cdot (\partial_x \boldsymbol{m} \times \partial_y \boldsymbol{m})\, dxdy$ ($Q\pm1$) and helicity $\gamma$ ($\gamma = 0, \pm\pi/2$ and $\pi$) [2]. When an external vertical electric field $\varepsilon$ is applied on the multiferroics monolayer to alter the direction of electrical polarization **P** (Supplementary Fig. S1a) [31], the DMI also switches its sign and causes a consequence of skyrmion diameter *d* variation (Supplementary Fig. S1b). Accordingly, the helicity of skyrmion switches from $\gamma = 0$ ($\pi$) to $\gamma = \pi$ (0) within a period T = 400 ps as shown in Supplementary Fig. S1c.

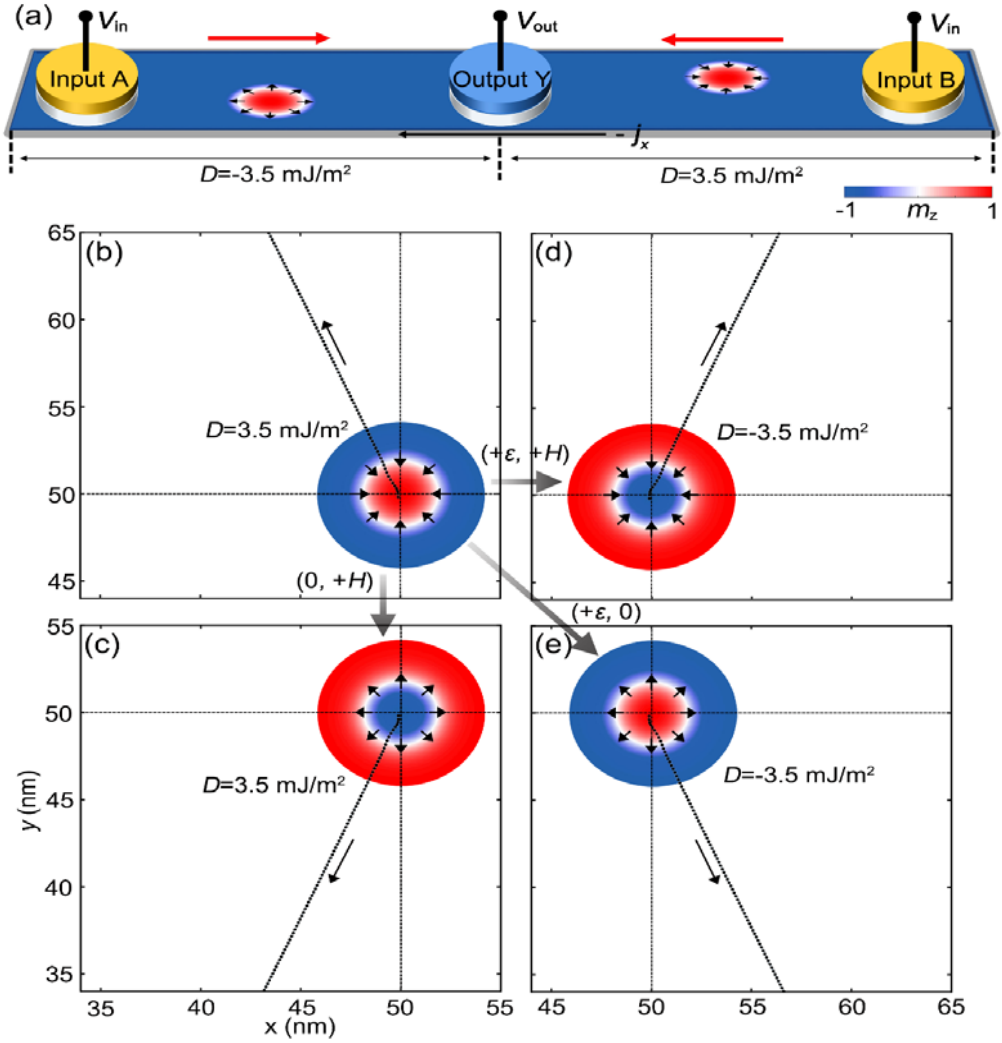

**Figure 1.** Schematic of the logic device and the evolution of four skyrmions motion driven by the same in-plane current. (a) Skyrmions-based single-nanotrack logic device with two inputs at both ends and one output in the middle. In the middle area of the nanostrip, a chirality barrier is formed by DMI with opposite sign. The skyrmions in (b), (c), (d) and (e) forms four different dynamic modes starting from the center of magnetoelectric multiferroics with a side length of 100 nm. The skyrmions in (c), (d), (e) can be obtained from that in (b) by applying corresponding electric field $\varepsilon$ or magnetic field $H$. The oblique dotted lines are their respective trajectories driven by the same current. The color scale indicates the out-of-plane component of magnetization, and has been used throughout this paper.

Since there are four different skyrmion states described by the chirality and polarity, a current applied in the same direction can induce four different skyrmion motion modes as shown in Fig. 1b-e. The topological charge sign switching induced by a magnetic field can only lead to the reversal of a transverse velocity (perpendicular to the driving current direction) [37], however, the chirality switching induced by an electric field can lead to the reversal of both transverse and horizontal velocities (parallel or antiparallel to the driving current direction). This mechanism can be demonstrated by a modified Thiele equation derived from the LLG equation by considering an isolated skyrmion as a rigid point-like particle [38-40] (see Methods and Supplementary Materials). This allows us to explore the interaction of magnetic skyrmions with different chirality in one single nanotrack driven by the same current as shown in Fig. 1a.

**Skyrmion – skyrmion pair annihilation, pinning and fusion**

To explore the interaction between skyrmions with different chirality, we designed a nanotrack comprising two parts with opposite DMI chirality separated by a constructed hereby non-volatile DMI chirality barrier as shown in the Fig. 2a. On both sides of the nanotrack with opposite DMI, two skyrmions with different helicity ($\gamma = 0$ and $\gamma = \pi$) are excited simultaneously. The two skyrmions will move towards to each other driven by a current via spin-orbit torque (SOT) and eventually collide in the middle area of the strip. Similar as two particles with positive and negative charges, the two skyrmions with opposite chirality can also attract each other. Since the pinning force of chirality barrier on the two skyrmions is less than the attractive force between them, they finally annihilate on the upper and lower sides of the track within 0.88 ns as shown in Fig. 2b (left panel). However, when two skyrmions cannot reach the DMI chirality barrier at the same time, the first skyrmion will be blocked by the barrier until another skyrmion with opposite chirality arrives, and then a similar skyrmion–skyrmion pair annihilation process occurs (Fig. 2b; Supplementary Video S1, 1.3 ns-1.7 ns). The pinning effect on skyrmion due to DMI chirality barrier can be observed more clearly in the Fig. 2b (right panel) and Supplementary Video S1 (1.7 ns-2.7 ns).

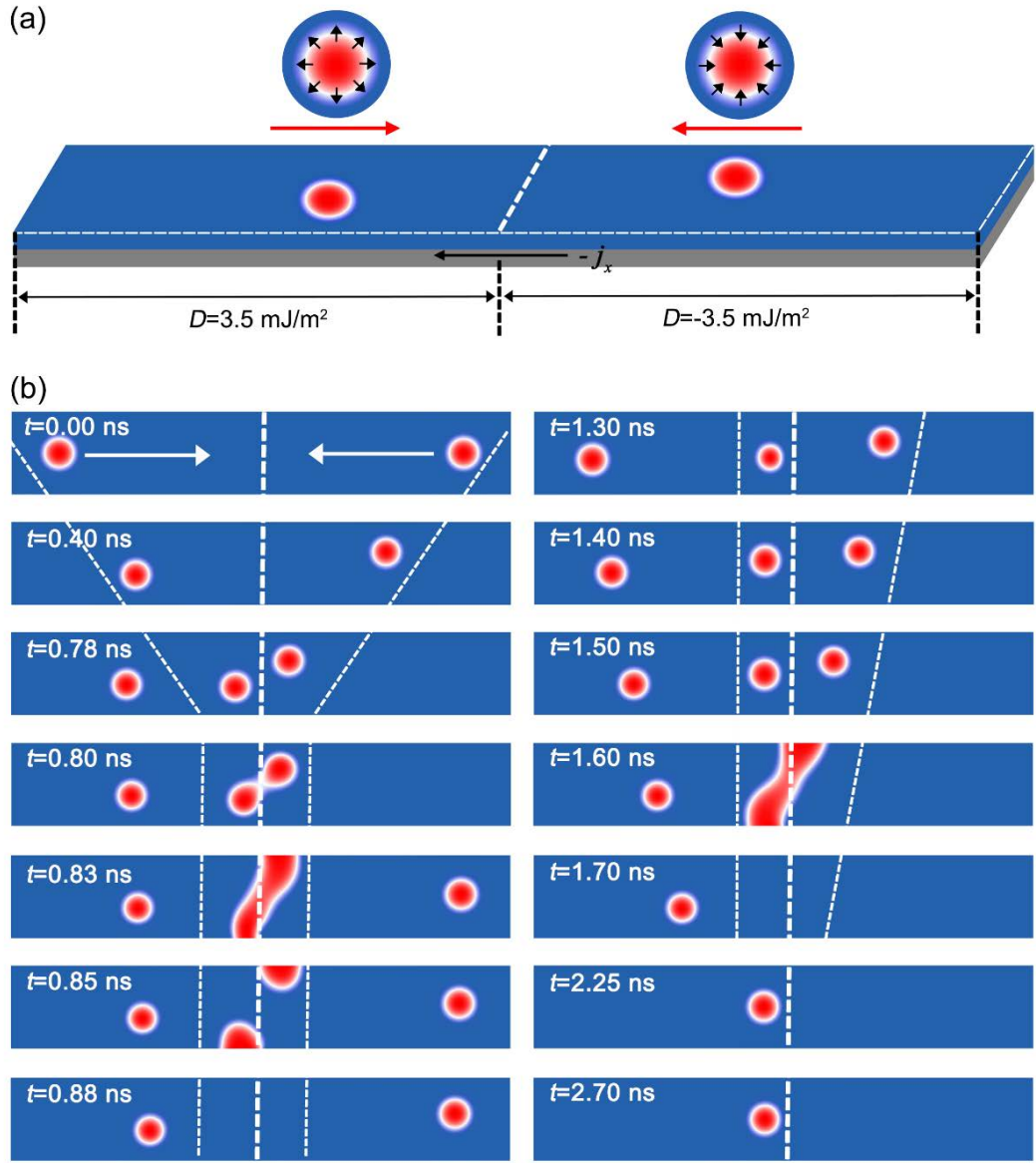

**Figure 2.** Simulation of skyrmion – skyrmion pair annihilation and pinning induced by DMI chirality barrier. (a) Scheme of the chirality barrier formed by nanotrack with opposite DMI chirality on the left and right. This structure can be realized by applying electric field pulse to the half of the nanotrack. (b) Snapshots of skyrmion–skyrmion pair annihilation and pinning process at different stages. The white dashed line in the middle of nanotrack indicates the DMI chirality barrier. The tilted dashed lines describe the transport process of magnetic skyrmions.

Fig. 3a shows the schematic diagram of double DMI chirality barriers. A positive electric field pulse can switch the DMI chirality of the brown area to positive and form chirality barriers in the middle of the strip, while a negative one can eliminate the chirality barriers. Compared to the chirality barrier in Fig. 2a, the double barriers in Fig. 3 also has a pinning effect on a single magnetic skyrmion, but the dynamics of skyrmions with opposite chirality will depend on both the attractive and pinning forces. When the double barriers are too narrow (below 6 nm) and the attractive force is stronger than the pinning one, the magnetic skyrmions will attract each other and annihilate together near the barrier. When the two magnetic skyrmions with opposite chirality approach to each other in double barriers wider than 6 nm but in the skyrmions interaction range, the pinning force will balance the attractive force, forming a non-trivial particle with topological charge Q=1.87 (Fig. 3b; Supplementary Fig. S2). The topological charge Q will be affected by the width of barrier. When the barrier is narrower, the magnetic skyrmions with opposite chirality will fuse more closely, and the topological charge will be close to 1, otherwise it will be close to 2.

Due to the strong magnetic anisotropy in CrN monolayer, the driving current density used in our logic operation is 28 MA/cm$^2$, which is nearly twice the current density used by other logic devices based on magnetic skyrmions [19]. A larger current density can further increase the attractive force between magnetic skyrmions with opposite chirality, and accelerate their annihilation or conversion process. However, an excessive current will annihilate them near the chirality barrier as shown in Supplementary Fig. S3. The Curie temperature of CrN is 805 K [41]. The simulated results have proved that neither thermal disturbances at room temperature nor the internal defects can affect the skyrmions dynamics. Moreover, it is shown that the dynamics of pinning, annihilation and fusion of skyrmions will not be affected in the strip with 12 nm transition region as shown in Supplementary Videos S2 (S3) for single (double) barrier. The magnetic skyrmion dynamics mentioned in this article are not limited to CrN monolayer, other structures with controllable DMI chirality can also realize similar skyrmion dynamics. The narrower chiral transition area and the larger skyrmion size can make the skyrmion dynamics more stable. The number and distribution of the DMI chirality barriers can be adjusted by electric field, which can realize the annihilation, pinning and fusion of skyrmion-skyrmion pairs with opposite chirality in a single nanotrack, thus providing a method for realizing skyrmions-based single-nanotrack logic gates.

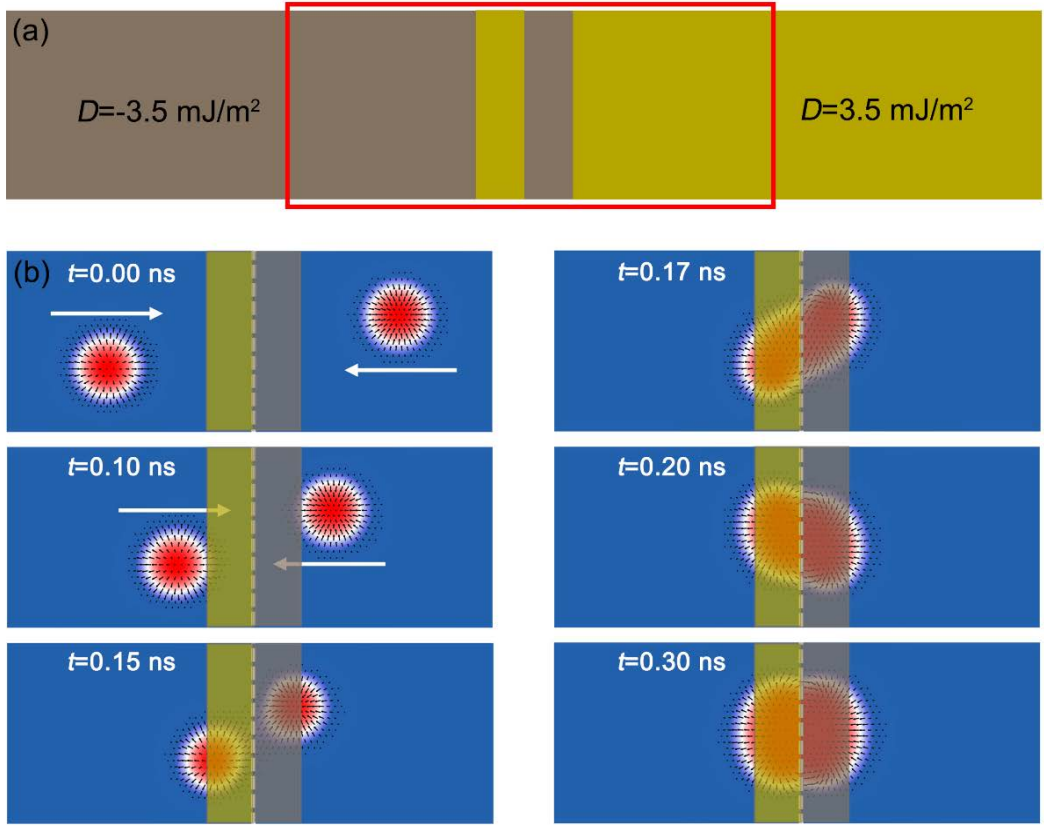

**Figure 3.** Skyrmion–skyrmion pair fusion induced by double DMI chirality barriers. (a) Schematic representation of double DMI chirality barriers formed by reversing the DMI chirality of the brown parts by electric field pulse. (b) Snapshots of the skyrmion–skyrmion pair fusion within the red box of (a) at different stages. The magnetization structure at $t$=0.30 ns is a non-trivial particle which is formed by the fusion of two skyrmions with opposite chirality.

**Reconfigurable single-nanotrack skyrmion logic gates**

Based on the proposed skyrmion–skyrmion pair annihilation and pinning effects, we can employ the electric field pulse to reconstruct non-volatile DMI chirality barrier in magnetoelectric multiferroics, and transform a single racetrack memory into any logic gates, realizing a complete logic family as displayed in Fig. 4. It is worth to note that here we realize one of the most challenging logic gates XOR in one single nanotrack, which is usually implemented by combining the AND and the OR gates. In the proposed structure shown in Fig. 4a, we simply reverse the DMI chirality of the left or right part of a monolayer racetrack memory to form a DMI chirality barrier (vacant bidirectional switch), and then successfully implement a logic XOR gate. For this logic application, two inputs are placed on the left and right sides, while an output is placed in the middle area (see red box area in Fig. 4a). Here we define the presence (resp. absence) of skyrmion corresponds to "1" (resp. "0") state. When only one of the two inputs is in "1" state, the skyrmion corresponding to this "1" state will be blocked in the red box area by the DMI chirality barrier resulting in the "1" output state. When both inputs are in "1" states, two skyrmions with different chirality meet and annihilate at the DMI chirality barrier yielding "0" state output. Thus, we realize the XOR gate function with only one single nanotrack (Fig. 4a; Supplementary Video S4).

Next, in the positive DMI region, we can use a downward electric field pulse to invert the DMI chirality to negative within a narrow area in the red output box (switch on position 1), and the XOR logic gate will be transformed into an OR gate. When two skyrmions with opposite chirality meet in the double barriers, they will not be annihilated, but will transform into a topological non-trivial particle in which two skyrmions with opposite chirality merge together as shown in Fig. 3b and Fig. 4b, then the output signal is "1" state (cf. also Supplementary state. To realize NAND gate, we turn on switch 2 to obtain wider DMI barriers as shown in Fig. 4c, and set two opposite chiral skyrmions in the middle area. Because the skyrmions with opposite chirality move backwards under the in-plane current, they will not meet and annihilate. In view of this, the input operations are similar to other logic gates, and the NAND function is hereby implemented (Fig. 4c; Supplementary Video S6). The difference between NAND and OR gates is the width of the DMI barriers, the latter is much thinner. In order to make the two skyrmions with opposite chirality attract each other and effectively fuse, our simulations show that the width should not be greater than 26 nm, which is the critical width for designing OR gate. When the barrier is wider than 26 nm, magnetic skyrmions with opposite chirality will be isolated on both sides of the barrier, and we can implement NAND logic operation.

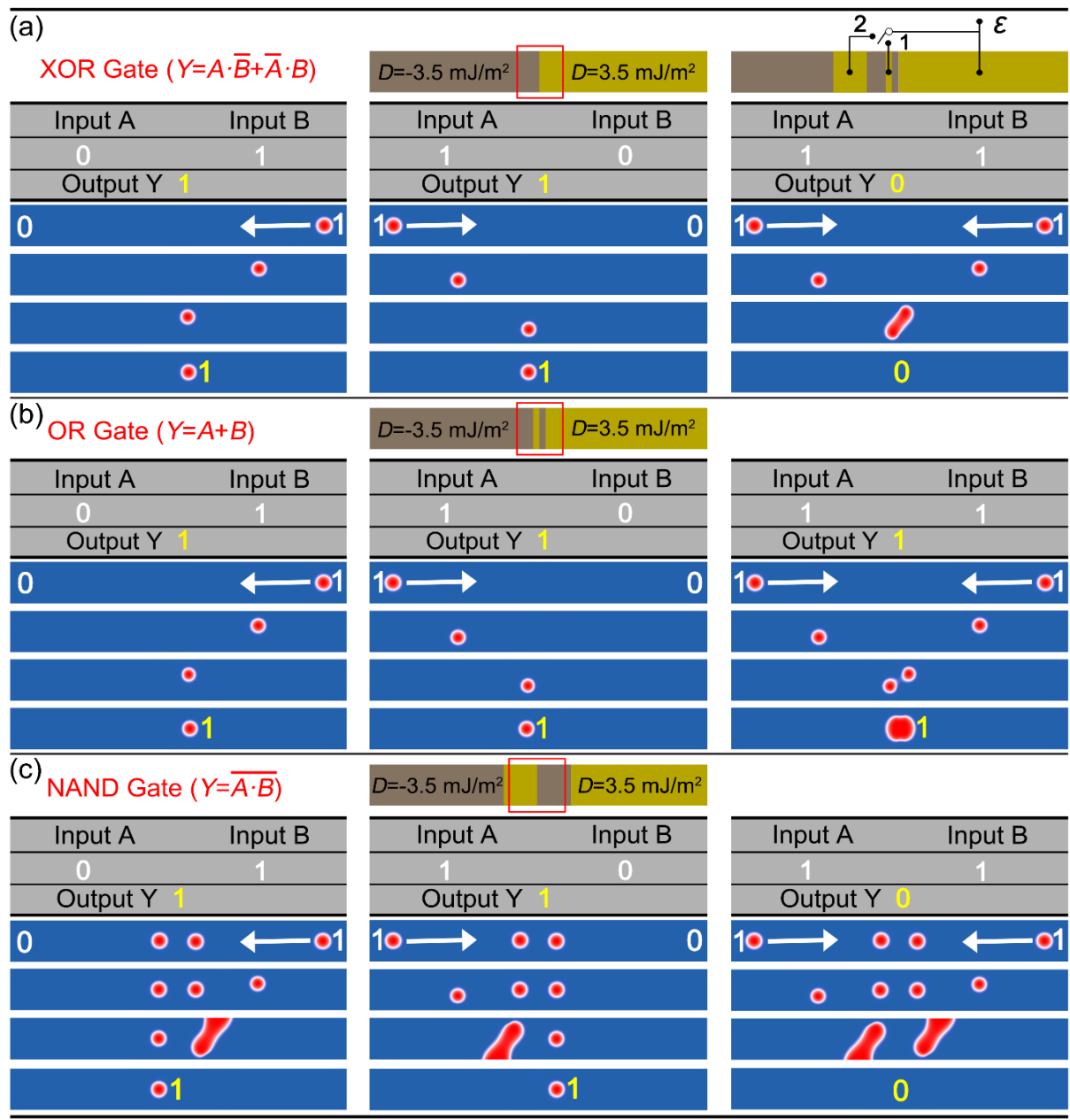

**Figure 4.** Reconfigurable single-nanotrack skyrmion logic gates. XOR/OR/NAND operations of the reconfigurable nanotrack with size 30 nm × 240 nm by turn the switch in (a) idle, (b) to 1 and (c) to 2, respectively. The electric field pulse can be removed after the DMI chirality is reversed, and the DMI chirality barrier will remain unchanged. XNOR/NOR/AND functions can be converted from above logic gates by switching the magnetization of fixed layer in magnetic tunnel junction (output-MTJ), and only half of the NAND gate can be used to realize the NOT gate.

Video S5). The logic NAND gate requires both inputs being in "0" state while the output in "1"

In the design of the above reconfigurable logic nanotrack, a magnetic tunnel junction (output-MTJ) is used to read skyrmions [42]. The bottom multiferroics adjacent to heavy metal (HM) layer works as the free layer of the MTJ. The top ferromagnetic (FM) layer acts as the fixed layer (Supplementary Fig. S4). When a skyrmion reaches bottom layer of the output-MTJ under driving current, the polarity direction of the skyrmion will be opposite to the magnetization of the fixed layer, and the output-MTJ will be in a high-resistance state resulting in a high voltage. If the magnetization of the fixed layer is switched, the output-MTJ will be in a low-resistance state when a skyrmion is detected, giving a low voltage as an output. In addition, when the MTJ is in a low-resistance state, it can also realize the reconstruction of logic gates in the chirality switching of multiferroics. Therefore, based on the XOR, OR, and NAND designed above, XNOR, NOR and AND functions can be similarly realized by switching the magnetization of MTJ. Using only one of the left and right parts of the NAND gate, it is also possible to implement the NOT gate [11].

It is worth noting that after each logic operation, the logic gate needs to be restored. The usual practice is to increase significantly the driving current in order to annihilate topological magnetic particles as shown in the simulation results in Supplementary Fig. S3. Alternatively, here we reintroduce (re-input) a skyrmion with opposite chirality in the opposite direction to perform the so-called erasing (reset) operation, which is more energy-efficient but needs the feedback from the previous logic operation. The detailed process is displayed in the corresponding video for each logic gate. In the initial state for NAND operation, we need to set two magnetic skyrmions in the middle of the nanotrack. The chirality of the two magnetic skyrmions is opposite. The two skyrmions can be introduced through MTJs at the two ends of the nanotrack, and then transported to the target position by current to perform logic operations and output information in the middle of the nanotrack. The operating speed of the logic device depends on the moving speed of magnetic skyrmion(s). Shown by the simulation of skyrmion dynamics in Fig. 2, the time required for each logic operation is about 1 ns. The logic operations in this study are completely implemented in a nanotrack, and thus the logic device can be connected to a skyrmions-based memory, thereby integrating logic and storage functions.

Considering the applications of energy barrier in manipulating magnetic skyrmion dynamics [19,23,43-45], the pinning and depinning effects of DMI chirality barrier on magnetic skyrmions can also be employed in other devices, e.g. design of transistors to switch on/off circuits (Supplementary Fig. S5a and Video S7). It is worth to note that the magnetic skyrmions can be recycled after each operation to reduce energy consumption as shown in Supplementary Fig. S6. In practical applications of skyrmion racetrack memory, the speed of skyrmion motion will be affected by temperature and material defects, resulting in possible bit misalignment. Here, we propose to adjust the position of skyrmion by constructing DMI chirality barriers as shown in Supplementary Fig. S5b, where the bits can be reset and the reading error of the skyrmion can be avoid. The barriers used in our logic devices have chiral characteristic, which are different from the other ones such as magnetic anisotropy barriers [44]. Due to the DMI signs in both sides being opposite, magnetic skyrmions will not cross this barrier unless the chirality of the magnetic skyrmion is reversed, and the chirality barrier can quickly brake skyrmion even under a large driving current.

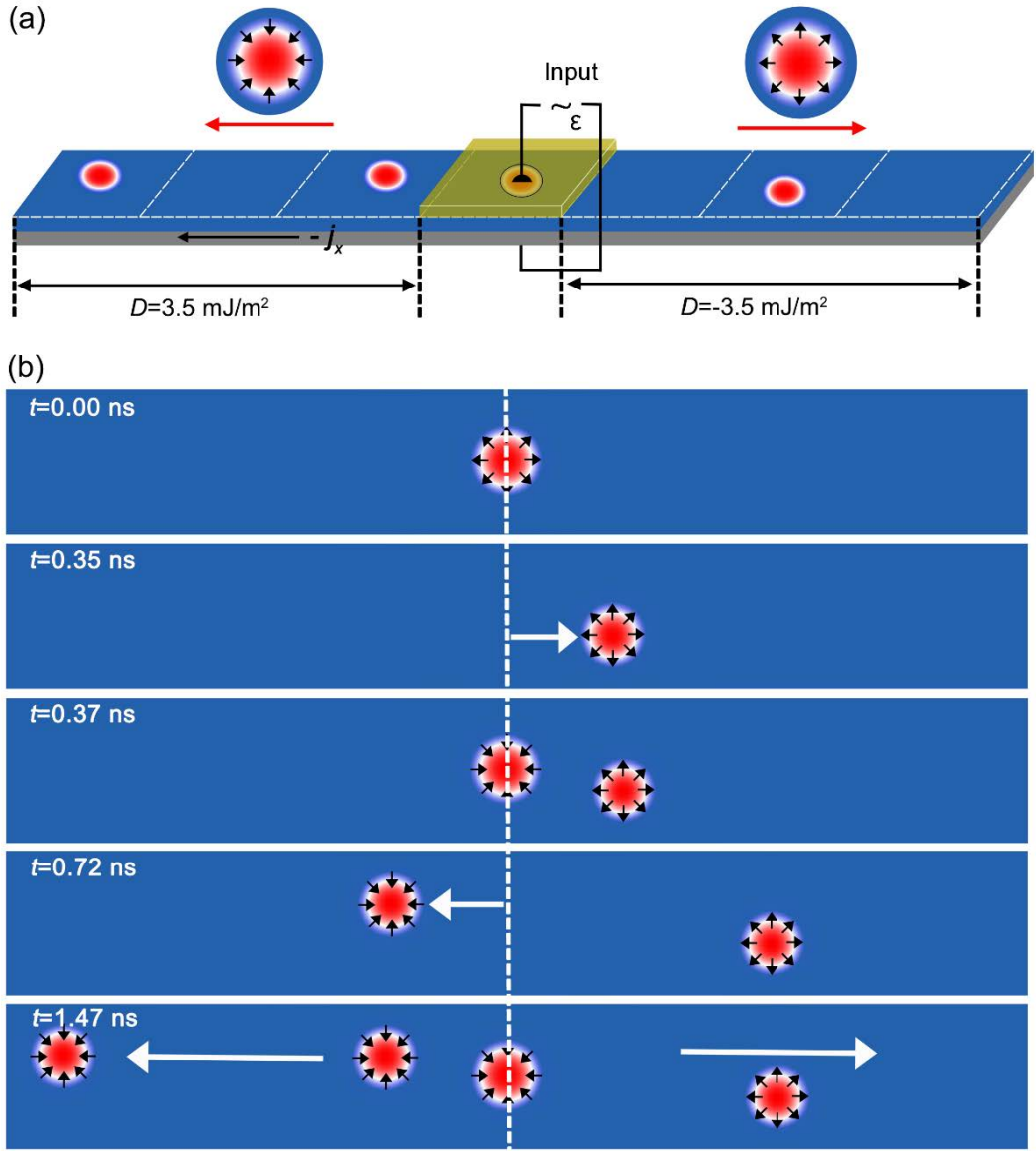

**Figure 5.** Shunting mode of skyrmions with opposite chirality. (a) Schematic diagram of the nanotrack with skyrmions shunting function. The DMI chirality in the middle region of the strip can be switched by electric field pulse to excite different chiral skyrmions. The DMI chirality in the left and right parts is opposite. (b) The shunting process of skyrmions in the nanotrack. Skyrmion with $\gamma = \pi$ moves to the left, and skyrmion with $\gamma = 0$ moves to the right.

**Skyrmions shunting**

The spin-polarized current can create isolated skyrmion in a nanotrack, and the DMI chirality of magnetic film determines the chirality of skyrmion. The skyrmions with opposite chirality can be driven in different directions by the same current via SOT [46,47]. This gives a way to the realization of skyrmions shunting in different states by switching the DMI chirality with an electric field pulse. Using a single nanotrack as shown in Fig. 5a, an applied electric field can control the chirality of skyrmion excited by the spin-polarized current in the middle area of the nanotrack. When the electric field pulse points downward, the DMI chirality of the middle area matches that of the right area, causing the skyrmion with helicity $\gamma = 0$ to move to the right driven by an in-plane current (Fig. 5b, 0 ns-0.35 ns). However, when the direction of the electric field pulse points upward, the DMI chirality in the middle area becomes the same as that of the left area, causing the excited skyrmion with opposite helicity ($\gamma = \pi$) moving to left (Fig. 5b, 0.37 ns-0.72 ns).

Using magnetic skyrmions or other non-trivial particles with different topological properties to encode bits is an attractive direction for possible practical applications [48-50]. Here, we use skyrmions with opposite chirality as information carriers to improve the robustness of data. Folding the nanotrack (Supplementary Fig. S7) from the middle results in the left (resp. right) half being on the upper (resp. lower) level. There will be only one skyrmion in the upper or lower channel, i.e. if a skyrmion exists as a bit in the upper (lower) nanotrack, there will be no skyrmion in the lower (upper) nanotrack. In addition, the skyrmions in the upper or lower channel will move to opposite directions under the same current. This effect will lead to that an excited skyrmion in the middle area can only choose one direction to move. Once the direction is determined, the corresponding area in the other direction will show a vacant state, thus a so-called natural complementary racetrack memory is constructed.

**CONCLUSION**

We demonstrate that the skyrmion chirality switching, pinning, pairwise annihilation, fusion and shunting can be achieved and switched by electric field pulse in magnetoelectric multiferroics. By locally controlling the DMI chirality, we can reconstruct the non-volatile DMI chirality barrier to switch the above-mentioned various magnetic skyrmion dynamic phenomena, allowing the implementation and reconfiguration of logic functions including AND, OR, NOT, NAND, NOR, XOR, and XNOR. Compared to other reconfigurable logic gates requiring the combination of multiple strips or the cascade of simple functions to perform two or more logic operations, we realize the implementation of the complete logic functions into one single nanotrack, thereby further simplifying the design of the spin-based logic device. Moreover, we also realize the "on" and "off" states of skyrmion transistor, and skyrmion bit reset via the pinning and depinning function of non-volatile DMI chirality barriers. Any two of these above functions or operations can be easily transformed from one to another by switching the sign of DMI chirality barrier, and skyrmions can be recycled after each operation. The study paves the way towards energy-efficient, high-density spin-based logic devices, illustrating the potential of using skyrmions for logic-in-memory applications.

**METHODS**

To investigate the dynamics of skyrmions, we performed finite-difference simulations using the Landau-Lifshitz-Gilbert (LLG) equation with a damping-like torque:

$$\frac{d\boldsymbol{m}}{dt} = -\frac{\gamma}{1+\alpha^2}(\boldsymbol{m} \times \mu_0 \boldsymbol{H}_{\text{eff}}) - \frac{\alpha\gamma}{1+\alpha^2}[\boldsymbol{m} \times (\boldsymbol{m} \times \mu_0 \boldsymbol{H}_{\text{eff}})] + \frac{u}{a}(\boldsymbol{m} \times \boldsymbol{p} \times \boldsymbol{m}),$$

where $\boldsymbol{m}$ indicates the magnetization unit vector, $\boldsymbol{H}_{\text{eff}}$ is the effective field which is defined as $\boldsymbol{H}_{\text{eff}} = -\mu_0^{-1}\frac{\partial E}{\partial \boldsymbol{m}}$ with following terms: exchange field $\boldsymbol{H}_{\text{exch}}$, perpendicular magnetic anisotropy (PMA) field $\boldsymbol{H}_{\text{anis}}$, demagnetization field $\boldsymbol{H}_{\text{demag}}$, and DMI field $\boldsymbol{H}_{\text{DMI}}$. $\boldsymbol{p}$ stands for the spin polarization unit direction, $\gamma$ and $\alpha$ represent the gyromagnetic ratio and the Gilbert damping coefficient, respectively. $u = |\frac{\gamma\hbar}{\mu_0 e}|\frac{j\theta_{\text{SH}}}{2M_s}$ indicates the spin torque coefficient with $j$ being the current flowing in the heavy metal (HM) layer. $\theta_{\text{SH}}$, $\hbar$ and $e$ indicate the spin Hall angle, the reduced Planck constant and the electron charge, respectively.

In the present study, material parameters for CrN monolayer are chosen based on first-principles calculations in the absence of electric field: exchange constant A=13.2 pJ/m, saturation magnetization $M_s$=1289 KA/m, magnetocrystalline anisotropy $K_{uz}$=1992 KJ/m$^3$, DMI energy constant $D=\pm 3.5$ mJ/m$^2$. The calculated material parameters with out-of-plane electric field $\varepsilon$ sweeps from -0.5 V/Å to 0.5 V/Å can be referred to Ref. [31]. We performed micromagnetic simulations in a sample of 100 nm×100 nm, and the grid was set to a quadrilateral with side length $a$=1 nm, which is sufficiently small compared to the typical exchange length and the 10 nm skyrmion. In addition, the width of logic device is 30 nm, and the length is 8 times of the width. Our simulation results indicate that the width of nanotrack should be less than 40 nm, so that the distance between the magnetic skyrmions with opposite chirality is small enough to interact. The main function of current in this study is to drive magnetic skyrmions towards the middle area of nanotrack for logic operations, and the driving current density should be above 10 MA/cm$^2$ to drive effectively skyrmions. In order to prevent skyrmions from annihilating at the boundary of nanotrack, however, the driving current density should not exceed 35 MA/cm$^2$.

It is worth to note that the construction of DMI chirality barrier and the resulting physical conclusions can be realized in structures of multilayer or 2D magnetoelectric multiferroics in which the DMI chirality is electrically switchable, which is not limited to CrN monolayer. When the magnetic parameters (exchange interaction, perpendicular magnetic anisotropy and DMI) meet the generation conditions for magnetic skyrmions, other structures or materials with reduced switchable DMI can also be employed for the implementation and realization of logical operations. In order to verify that the logic device still works with a reduced DMI value, we decrease the intensity of PMA and DMI to $K_{uz} = 1.7 \times 10^5$ J/m$^3$ and $D = \pm 1$ mJ/m$^2$, respectively, to maintain the stable existence of magnetic skyrmion. The simulation results show that the chirality barrier can still control the magnetic skyrmion dynamics including pinning, annihilation and fusion similar to those in Fig. 2 and Fig. 3, and have no influence on the implementation of logic operations.

**FUNDING**

This work was supported by "Pioneer" and "Leading Goose" R\&D Program of Zhejiang Province under Grant No. 2022C01053; National Natural Science Foundation of China (Grant Nos. 11874059 and 12174405); European Union's Horizon 2020 research and innovation Programme under grant agreement 881603 (Graphene Flagship); Ningbo Key Scientific and Technological Project (Grant No. 2021000215); Zhejiang Provincial Natural Science Foundation (Grant No. LR19A040002); Beijing National Laboratory for Condensed Matter Physics (No.2021000123); and China Postdoctoral Science Foundation (2021M703314).

**AUTHOR CONTRIBUTIONS**

H.Y. proposed and supervised the project. D.Y. performed the micromagnetic simulations. D.Y., H.Y., M.C. and A.F. co-wrote the manuscript. All authors discussed the results and contributed to the manuscript.

*Conflict of interest statement.* None declared.

# Supplementary Materials for

# Skyrmions-based logic gates in one single nanotrack completely reconstructed via chirality barrier

Dongxing Yu[1], Hongxin Yang[1,4,*], Mairbek Chshiev[2,5], Albert Fert[3]

[1]Quantum Functional Materials Laboratory, Ningbo Institute of Materials Technology and Engineering, Chinese Academy of Sciences, Ningbo 315201, China

[2]Université Grenoble Alpes, CEA, CNRS, Spintec, 38000 Grenoble, France

[3]Unité Mixte de Physique, CNRS, Thales, Université Paris-Sud, Université Paris-Saclay, Palaiseau 91767, France

[4]Center of Materials Science and Optoelectronics Engineering, University of Chinese Academy of Sciences, Beijing 100049, China

[5]Institut Universitaire de France (IUF), 75231, Paris, France

*Email: hongxin.yang@nimte.ac.cn

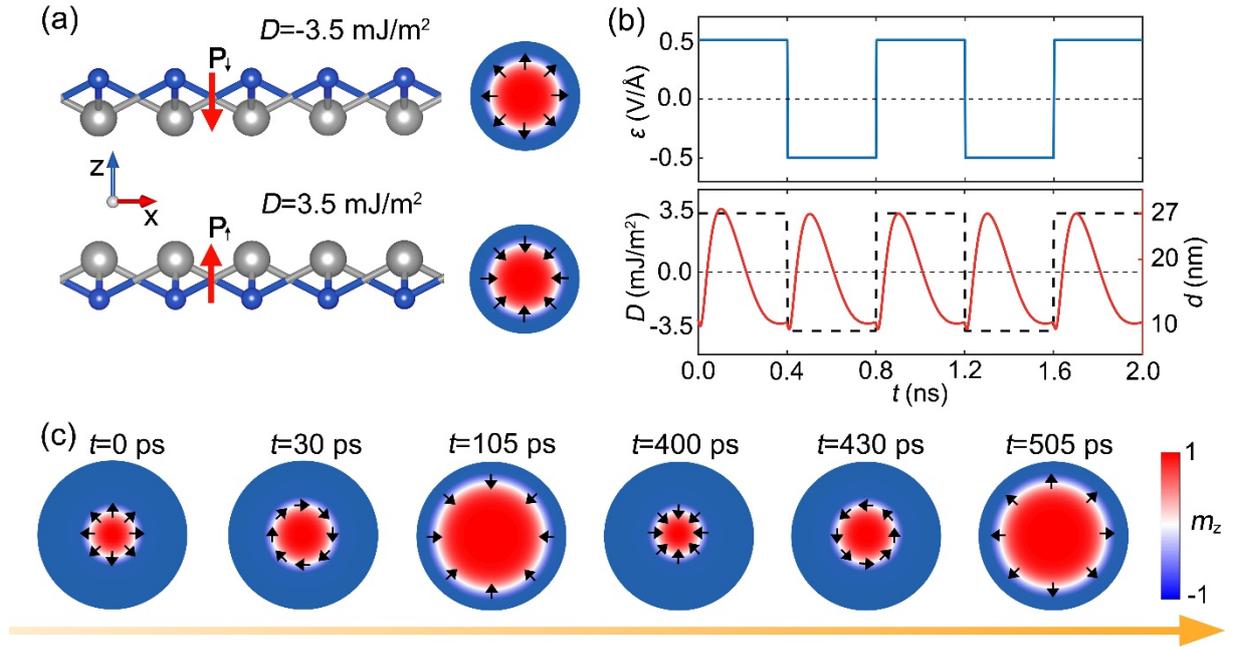

**Figure S1.** Simulated skyrmion chirality switching accompanied by breathing mode. (a) Two-dimensional multiferroics monolayer with controllable fluctuations. The direction of Rashba-type DMI and the chirality of skyrmions can be switched by electric field pulse via electrical polarization **P**. (b) Evolution of the skyrmion diameter $d$ and DMI constant $D$ (down) driven by electric field $\varepsilon$ (up) as a function of time. Snapshot images in (c) illustrating the skyrmion helicity ($\gamma = 0 \rightarrow \gamma = -\frac{\pi}{2} \rightarrow \gamma = \pi \rightarrow \gamma = \frac{\pi}{2} \rightarrow \gamma = 0$) and size at different stages. With the application of periodic electric field, the subsequent skyrmion dynamic (0.8 ns – 2 ns) also shows a good stability.

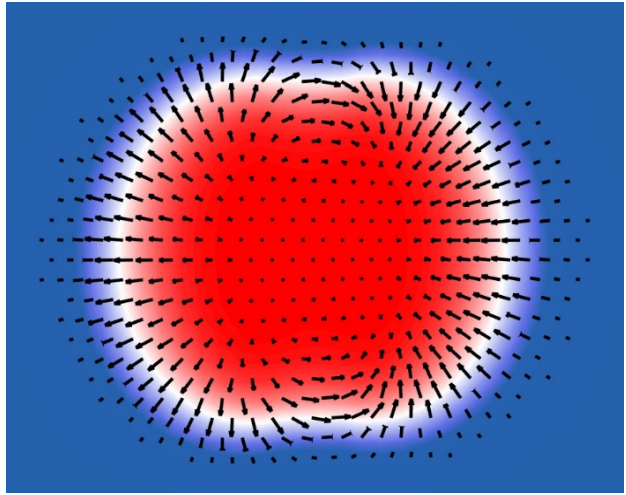

**Figure S2.** Topological non-trivial particle in the OR operation. The magnetization configurations with topological charge Q=1.87, in which two skyrmions with opposite chirality merged together. The merging is the result of the pinning effect of DMI chirality barrier and the mutual attraction of magnetic skyrmions with opposite chirality.

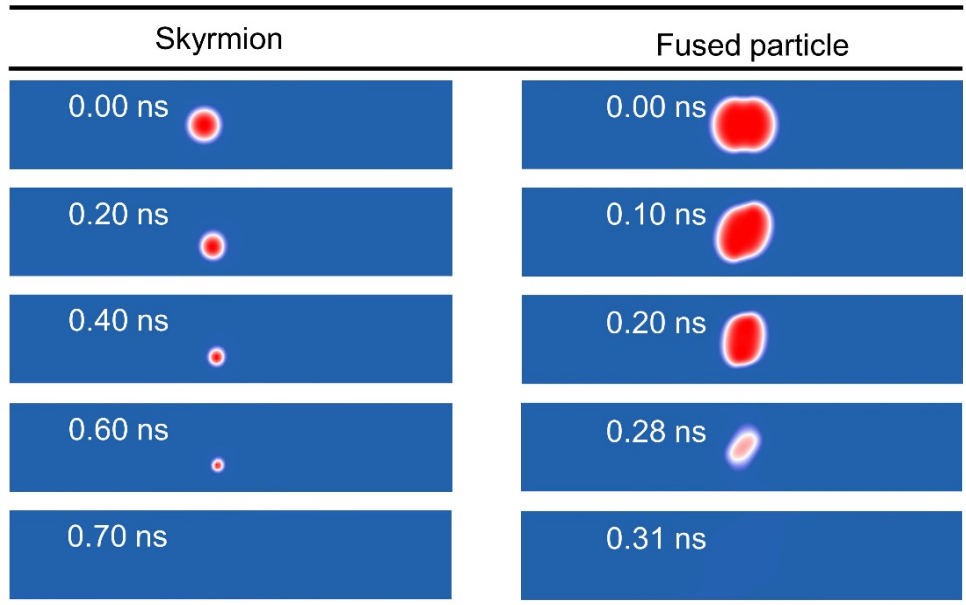

**Figure S3.** Snapshots of simulation at different times for magnetic skyrmion (left panel) and fused non-trivial particle (right panel) erased by a large current density 45 MA/cm$^2$.

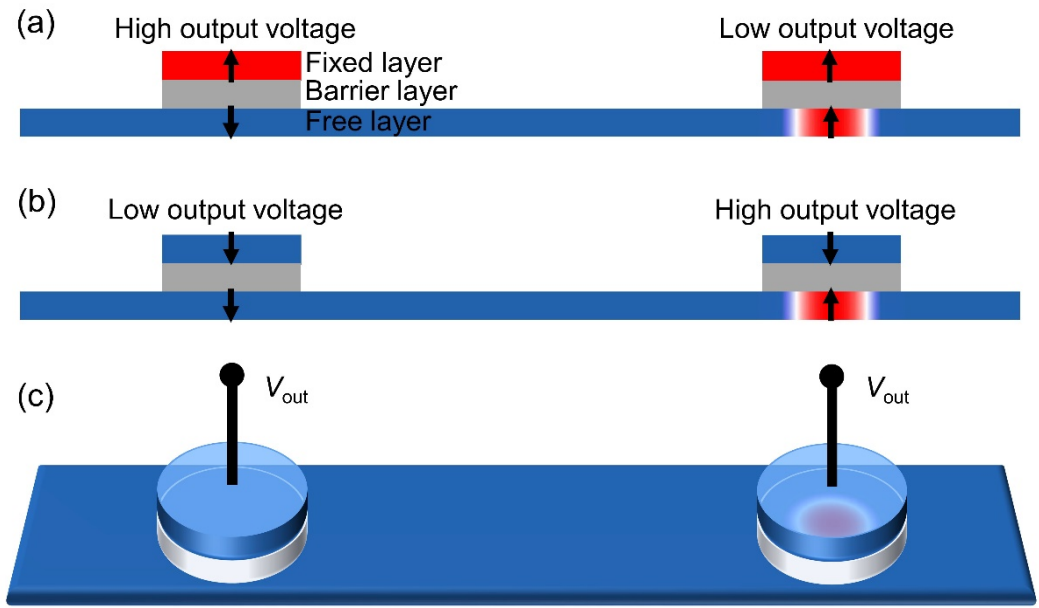

**Figure S4.** Magnetic tunnel junction (output-MTJ) structure. Four different magneto-resistive states and the corresponding output voltages for the fixed layer magnetization pointing (a) up, (b) down. (c) Schematic diagram of detecting the background state (left output-MTJ) and the magnetic skyrmion state (right output-MTJ) in a racetrack.

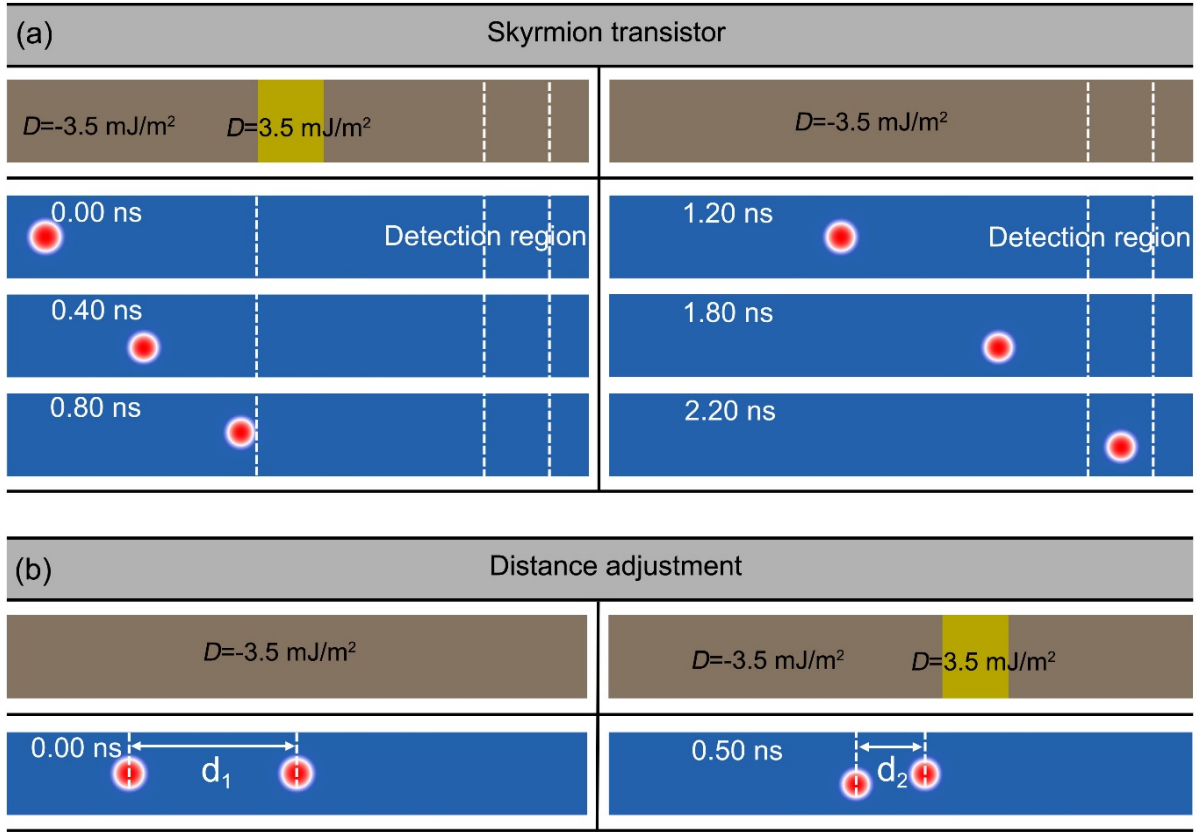

**Figure S5.** Device structures with size 30 nm × 210 nm for skyrmion transistor and skyrmion bit reset. (a) "off" states (left column) and "on" states (right column) of a skyrmion transistor based on the pinning/depinning function of DMI chirality barrier (the brown region in the strip) controlled by electric field pulse. (b) Distance adjustment between skyrmions. The DMI chirality barrier in the right panel pins the right skyrmion. The current drives the left skyrmion to move to the right, shortening the distance between the two skyrmons from d1 to d2.

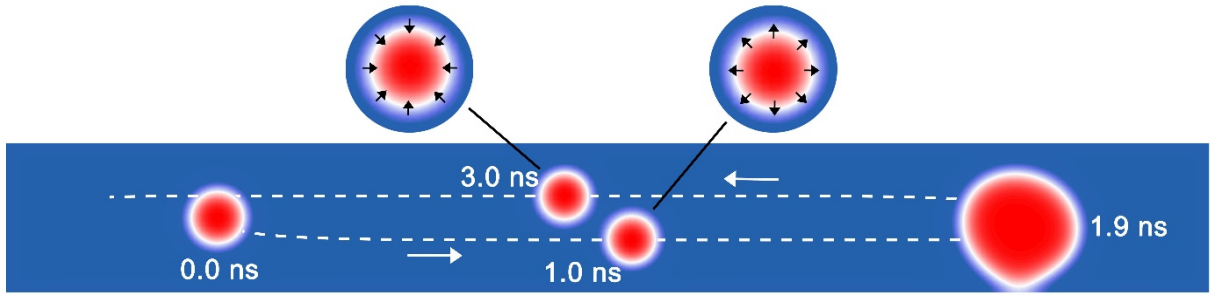

**Figure S6.** Back and forth motion of skyrmion (skyrmion recycling) switched by electric pulse without changing the in-plane current direction.

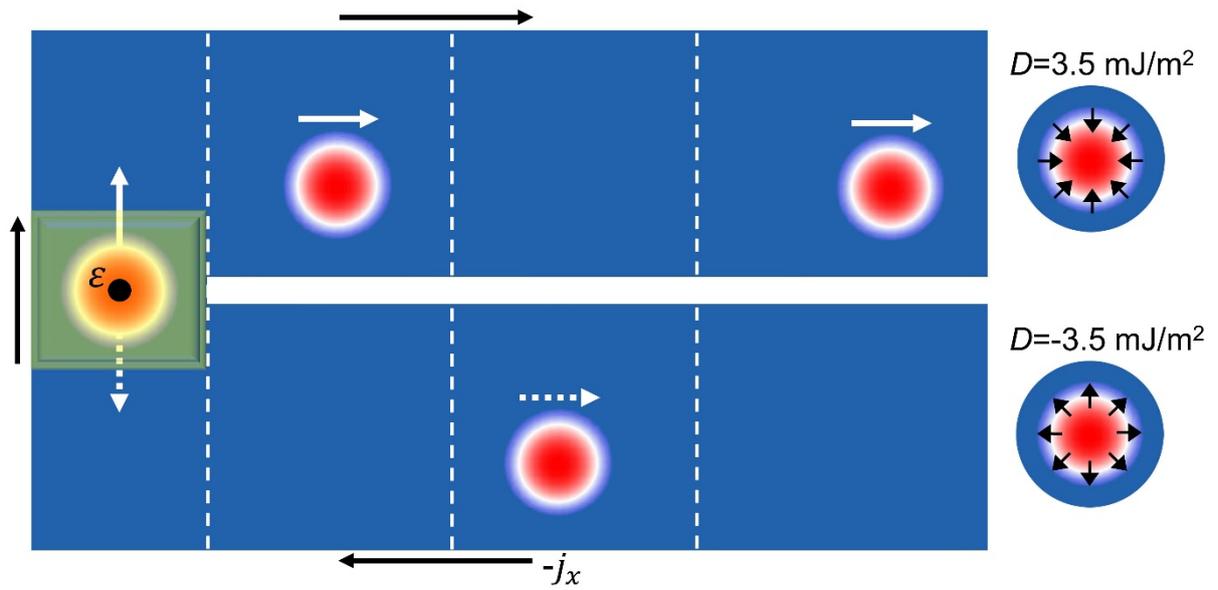

**Figure S7.** Complementary racetrack memory. This structure can be constructed by folding the nanotrack in Fig. 4. DMI chirality of the upper and lower channels is opposite, and the vertical electric field $\varepsilon$ can control the chirality of the skyrmions excited on the left. When the electric field is positive, the newly nucleated skyrmion will move to the upper channel, and vice versa.

## Motion of magnetic skyrmions with opposite chirality driven by a same current

To investigate the effect of the skyrmion chirality switching on its motion mode, the Thiele's equation derived from the LLG equation is employed:

$$\boldsymbol{G} \times \boldsymbol{v} - \alpha \boldsymbol{\mathcal{D}} \cdot \boldsymbol{v} + \boldsymbol{F}_{\text{SOT}} = 0.$$

Here, the first term represents the Magnus force that gives rise to the Skyrmion Hall Effect (SkHE) [1-5], gyrovector $\boldsymbol{G} = (M_s h/\gamma) 4\pi Q \boldsymbol{z}$ with topological (or skyrmion) number $Q = 1/(4\pi) \int \boldsymbol{m} \cdot (\partial_x \boldsymbol{m} \times \partial_y \boldsymbol{m}) dx dy$, and $\boldsymbol{v}$ being the skyrmion velocity. $\alpha$ indicates the magnetic damping coefficient and $\boldsymbol{\mathcal{D}}$ is the dissipative tensor. The force arising from spin-orbit torque (SOT) $F_{\text{SOT}}$ can be identified as a volume integral for each component [6]:

$$F_{x,\text{SOT}} = -\int d\mathcal{V} \boldsymbol{B}_{\text{SOT}} \cdot \frac{\partial}{\partial x} \boldsymbol{M} = \tau_{\text{DL}} \boldsymbol{\zeta} \cdot \int d\mathcal{V} \boldsymbol{m} \times \frac{\partial}{\partial x} \boldsymbol{m},$$

where the spin polarization along $\boldsymbol{\zeta}$ is parallel to the y-axis for current flowing along x. Because the x component of the force (along the current) is the the same as the part of the interface DMI that which involves the x gradient, the force $F_{x,\text{SOT}}$ from SOT will on a skyrmion depends on the its chirality and, polarity of skyrmion besides in addition to the spin polarization $\boldsymbol{\zeta}$. This means that when either the skyrmion helicity or topological charge is changed separately, the sign of the force $F_{x,\text{SOT}}$ will also be altered, and the skyrmion motion can be reversed without changing the direction of the current. But when the two are reversed at the same time, the force $F_{x,\text{SOT}}$ remains unchanged, only the SkHE will be reversed by the magnus force. Therefore, for the skyrmion driven by a transverse current, the skyrmion switching of the chirality will cause the direction of the force $F_{x,\text{SOT}}$ to be reversed, i.e., giving the same effect as changing the current direction (the spin polarization $\boldsymbol{\zeta}$), thereby reversing the direction of skyrmion movement.